\documentclass[draft,jgrga]{agutex}

 \usepackage{lineno}
\usepackage{graphicx}
  \setkeys{Gin}{draft=false}
\authorrunninghead{POLUIANOV ET AL.}

\titlerunninghead{COSMOGENIC ISOTOPES IN THE ATMOSPHERE}

\authoraddr{Corresponding author: I. Usoskin
University of Oulu, Finland,
(ilya.usoskin@oulu.fi)}

\begin{document}

%
%

\title{Production of cosmogenic isotopes $^{7}$Be, $^{10}$Be, $^{14}$C, $^{22}$Na and $^{36}$Cl in the atmosphere:
 Altitudinal profiles of yield functions}
%
%

%
%



 \authors{S. Poluianov,\altaffilmark{1}
  G.A. Kovaltsov,\altaffilmark{2}
  A.L. Mishev,\altaffilmark{1}
  I.G. Usoskin,\altaffilmark{1,3}}

\altaffiltext{1}{ReSoLVE Center of Excellence, University of Oulu, Finland.}
\altaffiltext{2}{Ioffe Physical-Technical Institute of Russian Academy of Sciences, St. Petersburg, Russia.}
\altaffiltext{3}{Sodankyl\"a Geophysical Observatory (Oulu unit), University of Oulu, Finland.}

\noindent \textbf{Key points.}
\begin{itemize}
\item
A new consistent set of yield functions for cosmogenic isotopes 7Be, 10Be, 14C, 22Na, 36Cl by cosmic rays in the atmosphere is presented.
\item
For the first time, a detailed altitudinal profile of the production is given.
\item
The results can be straightforwardly used in the atmospheric chemistry and dynamics models.
\end{itemize}

\begin{abstract}
New consistent and precise computations of the production of five cosmogenic radio-isotopes,
 $^7$Be, $^{10}$Be, $^{14}$C, $^{22}$Na and $^{36}$Cl, in the Earth's atmosphere by cosmic rays are presented
 in the form of tabulated yield functions.
For the first time, a detailed set of the the altitude profiles of the production functions is provided
 which makes it possible to apply the results directly as input for atmospheric transport models.
Good agreement with most of the earlier published works for columnar and global isotopic production rates is shown.
Altitude profiles of the production are important, in particular for such tasks as studies of strong solar particle
 events in the past, precise reconstructions of solar activity on long-term scale,
 tracing air-mass dynamics using cosmogenic radio-isotopes, etc.
As an example, computations of the $^{10}$Be deposition flux in the polar region are shown for the last decades and also
 for a period around 780 AD and confronted with the actual measurements in Greenland and Antarctic ice cores.
\end{abstract}

\begin{article}

\section{Introduction}

Earth is permanently bombarded by high-energy nucleonic particles -- cosmic rays, which produce nucleonic-muon-electromagnetic
 cascades in the Earth's atmosphere.
As a sub-product of the cascade, radioactive isotopes can be produced, called cosmogenic nuclides.
Measurements of the abundance of long-living cosmogenic radionuclides in the atmosphere and terrestrial
 archives (ice cores, tree trunks, sediments, etc.) form a very important tool to study atmospheric processes and
 interaction between different reservoirs (see, e.g., books by \citet{dorman04} and \citet{beer12}).
This also offers a reliable quantitative method to study solar activity on the long time scale
 \citep{mccracken04,solanki_nat_04,vonmoos06,muscheler07,steinhilber12,inceoglu15,usoskin_LR_13,usoskin_AAL_14}.
Most important for these purposes are cosmogenic isotopes
 $^7$Be (half-life $\approx$53 days), $^{22}$Na (2.6 years), $^{14}$C (5730 years), $^{36}$Cl ($3\cdot10^5$ years)
  and $^{10}$Be ($1.4\cdot 10^6$ years), and many studies are based on these data.

Although the relation between cosmic ray variability and cosmogenic isotopes is qualitatively obvious, their
 quantitative modelling is difficult,
 since they are produced in complex atmospheric cascades which require extensive computations.
First numerical models of cosmogenic nuclide production were developed already in the 1960--1970s
 \citep[e.g.][]{lal62,lingenfelter63,obrien79}, using either direct modelling or (semi)empirical parameterizations.
A benchmark was achieved by \citet{masarik99} (updated as \citet{masarik09}) who applied modern high-performance
  computers for direct Monte-Carlo simulations of the atmospheric cascade to model production rates of isotopes
  $^7$Be, $^{10}$Be, $^{14}$C and $^{36}$Cl.
Unfortunately, their computations were made for a prescribed spectrum of cosmic rays without the yield-function approach
 (see Sect.~\ref{Sec:YF}).
This shortcoming was soon overcome in a number of original works presenting production yield functions for different
 isotopes:
\citet{webber03} and \citet{webber07} calculated yield functions for $^7$Be, $^{10}$Be and $^{36}$Cl using FLUKA Monte-Carlo
 code \citep{fasso01};
\citet{usoskin_7Be_08}, \citet{kovaltsov10} and \citet{leppanen12} calculated, using the CRAC (Cosmic Ray induced Atmospheric Cascade)
 model, based on CORSIKA Monte-Carlo tool \citep{heck98}, yield functions for cosmogenic $^7$Be, $^{10}$Be and $^{22}$Na, respectively;
 the yield function for production of $^{14}$C was calculated by \citet{kovaltsov12} using the GEANT4-based tool PLANETOCOSMICS \citep{desorgher09}.
Thus, it is presently a mixture of different yield functions calculated by different models with different assumptions
 and conditions.

For many tasks it is important to know detailed altitude profiles of the isotope production:
 studies of solar energetic particle events in the past \citep[e.g.,][]{usoskin_GRL_SCR06,webber07,miyake12,usoskin_775_13};
 detailed reconstructions of long-term solar activity using isotopes of $^{10}$Be and $^{36}$Cl including
  realistic atmospheric transport \citep[e.g.,][]{mccracken_JGR_04,field06,pedro06,heikkila09,delaygue15};
 in situ atmospheric measurements of $^{14}$C \citep{jockel99,jockel03,kanu16};
 tracing of air-mass dynamics and water flows using cosmogenic radio-isotopes
  \citep[e.g.,][]{jordan03,sakaguchi05,leppanen12,ioannidou15,pacini_AE_15}.
However, the previously published yield functions were presented either without altitudinal resolution,
 giving only atmospheric columnar or global production of isotopes, or with very rough altitudinal resolution,
 insufficient for detailed computations.
This made it difficult to solve the above tasks independently without involving
 modelling groups making additional laborious detailed simulations.
While \citet{masarik99} provided computations with a detailed vertical resolution,
  they were not based on the yield-function formalism, limited to a prescribed energy spectrum of GCR,
  thus being inapplicable to, e.g., an analysis of solar energetic particle events.
Moreover, they included $\alpha-$ and heavier particles as scaled protons, which is not exactly correct
 \citep{webber07}.

Here we present a new computation of yield functions for a set of widely used cosmogenic isotopes, viz.
 $^{7}$Be, $^{10}$Be, $^{14}$C, $^{22}$Na and $^{36}$Cl, in the Earth's atmosphere.
These isotopes are produced in the atmosphere in nuclear reactions on nitrogen, oxygen and argon, induced by
 nucleonic particles (neutrons, protons and $\alpha-$particles) of the nucleonic cascade.
We provide a set of new consistent computations of the yield function production for cosmogenic isotopes,
 using one and the same Monte-Carlo model (GEANT4) with fixed physical sub-model options, atmospheric parameters and basic assumptions,
 and assess uncertainties arising.
For the first time, we publish (see the Supporting information) a detailed set of altitude profiles of production functions
 of the cosmogenic isotopes making it possible for everyone to calculate the full 3D atmospheric production of isotopes.

\section{Yield function: Formalism and computational model}
\label{Sec:YF}

\subsection{Formalism}
A standard approach to model various effects of cosmic rays in the atmosphere,
 including production of cosmogenic isotopes, is based on the yield function formalism.

The yield function, $Y(E,h)$, is defined as the production (the number of atoms per gram of air)
 of the isotope, at given atmospheric depth $h$, by primary particles of type $i$ with the unit
 intensity (one primary particle with kinetic energy per nucleon $E$ in the interplanetary space per steradian and cm$^2$).
The units of $Y$ are [atoms g$^{-1}$ cm$^2$ sr].
The production rate $Q$ of cosmogenic isotope at time  $t$ is then defined as an integral of the product of the
 yield function and the energy spectrum of cosmic rays $J_i(E,t)$ [sr sec cm$^2$]$^{-1}$),
 above the energy $E_{\rm c}$ corresponding to the local geomagnetic rigidity cutoff $P_{\rm c}$:
\begin{equation}
Q(t,h,P_{\rm c}) = \sum_i\int_{E_{\rm c,i}}^\infty{Y_i(E,h)\cdot J_i(E,t)\cdot dE},
\label{Eq:Q}
\end{equation}
where the summation is over different types of primary cosmic ray particles (protons, $\alpha-$particles, etc.).
The relation between $E_{\rm c,i}$ and $P_{\rm c}$ (defined independently of the yield function computations) is
\begin{equation}
E_{{\rm c},i}=E_{\rm r}\cdot\left(\sqrt{1+\left({Z_i\cdot P_{\rm c}\over A_i\cdot E_{\rm r}}\right)^2\,}-1\right),
\label{Eq:Tc}
\end{equation}
where $Z_i$ and $A_i$ are the charge and mass numbers of particles, respectively, $E_{\rm r}=0.938$ GeV is the rest mass of a proton.
For computations of the yield function we considered, as primary particles, only protons and $\alpha-$particles.
Species heavier than helium can be effectively considered as scaled (by the nucleonic number) $\alpha-$particles
 \citep[see ][]{webber03}.
An advantage of this approach is that the production rate can be calculated for any type of the energy spectrum
 beyond the standard modulated spectrum of galactic cosmic rays (GCR),
 for example, for solar energetic particle events or hypothetical nearby supernova explosions.
Sometimes production of cosmogenic isotopes is calculated directly, without the yield
 function \citep{obrien79,masarik99,masarik09}, but this is related to a prescribed cosmic-ray spectrum and cannot be
 applied for a different or/and revised spectrum.

We note that the computational results are often given not as the strictly defined yield function $Y$ (see above) but
 the so-called `production function' $S$, which gives production of the cosmogenic isotope per one primary particle
 impinging on the top of the atmosphere \citep[e.g.,][]{webber07}.
Here we show and tabulate the production function $S$ which is a direct result of the simulations
 and which is related, for the isotropic flux of primary cosmic rays, to the true yield function as
\begin{equation}
Y=\pi\, S.
\label{Eq:Y}
\end{equation}
The factor $\pi$ appears as conversion between the flux on the top of the atmosphere and the CR intensity in the interplanetary space
\citep[cf., e.g.,][Chapter 1.6.2]{grieder01}.

\subsection{Numerical calculation of the yield function}

Simulations of the nucleonic cascade in the atmosphere were made using direct Monte-Carlo simulations by the
 general-purpose toolkit GEANT4 10.0 developed in CERN for modelling the particle transport and interactions \citep{Geant4-1, Geant4-2}.
For our task we applied the embedded physics list QGSP\_BIC\_HP (Quark-Gluon String model for high-energy interactions +
 Geant4 Binary Cascade + High-Precision neutron package) \citep{Geant4-Phys2013}.
The Earth's atmosphere was modelled in a realistic manner as a set of spherical layers with
 homogeneous properties according to the empirical model of the atmosphere NRLMSISE-00 \citep{picone02}.
The top of the model was set at the altitude of 100 km, and the layers had the thickness from 1 g/cm$^2$ (for the top 20 g/cm$^2$)
 to 10 g/cm$^2$ (for the ones below 20 g/cm$^2$).
The total depth of the atmosphere was set to 1050 g/cm$^2$, and the soil was not included into the model.

The primary cosmic rays were modelled, in each simulation, as the monoenergetic (viz. with a fixed energy) isotropic flux
 impinging on the top of the atmosphere.
We did a series of simulations for two types of cosmic rays, primary protons and $\alpha-$particles with fixed energies
 in the range from 20 MeV/nuc to 100 GeV/nuc with a quasi-logarithmically distributed values.
We note that the computations are presented not for energy bins but for the fixed energies as denoted in the
 supplement information Tables.
From these simulations, the depth-energy distributions of the fluxes of cascade particles (protons, neutrons and $\alpha-$particles,
 both primaries and secondaries) were stored as histograms with energy range 1 keV -- 100 GeV with
 logarithmic bins in energy (20 bins per decade).

We obtained sums of simulated secondary particles with their energy binned into energy bins of width $\Delta E'$ centered at
 the energy $E'$, that have crossed a given horizontal level (atmospheric depth, $h$), and applying a weight of $|\cos{\theta}|^{-1}$
 (where $\theta$ is the zenith angle of the secondaries) to account for the geometrical factor.
The minimum value of $|\cos{\theta}|$ was limited to 0.001 to avoid too high weights.
These sums were divided by the energy bin width $\Delta E'$ to correspond to the quantity $F_k(E',h)$ which is defined as
\begin{equation}
F_k(E',h)\equiv N_k(E',h)\, v_k(E'),
\end{equation}
where $N_k$ and $v_k$ are the concentration (in [MeV cm$^3$]$^{-1}$) and velocity of secondary particles of
 type $k$ with energy $E'$ at the atmospheric depth level $h$.
Then the production function $S$ (in units of atoms/g) at the given atmospheric level $h$ is defined as
\begin{equation}
S(E,h) = \sum_j \sum_k \int{\kappa_j\cdot F_{k}(E',h)\cdot \sigma_{j,k}(E')\cdot dE'}
\label{Eq:S}
\end{equation}
where $\kappa_j$ is the content of the target nuclei in one gram of air (atoms/g), $\sigma_{j,k}$ is the cross-section of the
 corresponding nuclear reactions, and summation is
 over the type of secondary particle $k$ and the type $j$ of the target nucleus.

Radiocarbon $^{14}$C is produced mostly via capture of secondary neutrons by atmospheric nitrogen which composes 78\%
 of the atmosphere by volume.
The $^7$Be and $^{10}$Be isotopes are produced by spallation of atmospheric nitrogen and oxygen (forming together
 about 99\% of the atmosphere by volume).
The $^{22}$Na and $^{36}$Cl isotopes are produced by spallation of atmospheric argon (about 1\% of the atmosphere by volume)
 which is much less abundant than nitrogen and oxygen.
In computations we adopted the cross sections from \citet{reyss81,lange95,jull98,webber03,tatischeff06,beer12} and also from
 the Experimental Nuclear Reaction Database (EXFOR/CSISRS) http://www.nndc.bnl.gov/exfor/exfor00.htm.
We note that cross-sections we used to compute the production of $^7$Be, $^{10}$Be, $^{14}$C and $^{22}$Na are the same as in our
 previous works \citep{usoskin_7Be_08,kovaltsov10,kovaltsov12,leppanen12}.
Transport of neutrons with energy below 1 keV, for production of $^{14}$C, was calculated in a way similar
 to the work of \citet{kovaltsov12}.

The number of simulated cascades was set to assure the statistical accuracy of the computed columnar isotope production
 to be better than 1\%.
It varied with the type of primaries and their initial energy, ranging from 1000 simulated cascades (for $\alpha-$particles with
 the energy of 100 GeV/nuc) to $2\cdot 10^7$ cascades (for 20 MeV protons).

\begin{figure*}
\centering \resizebox{14cm}{!}{\includegraphics{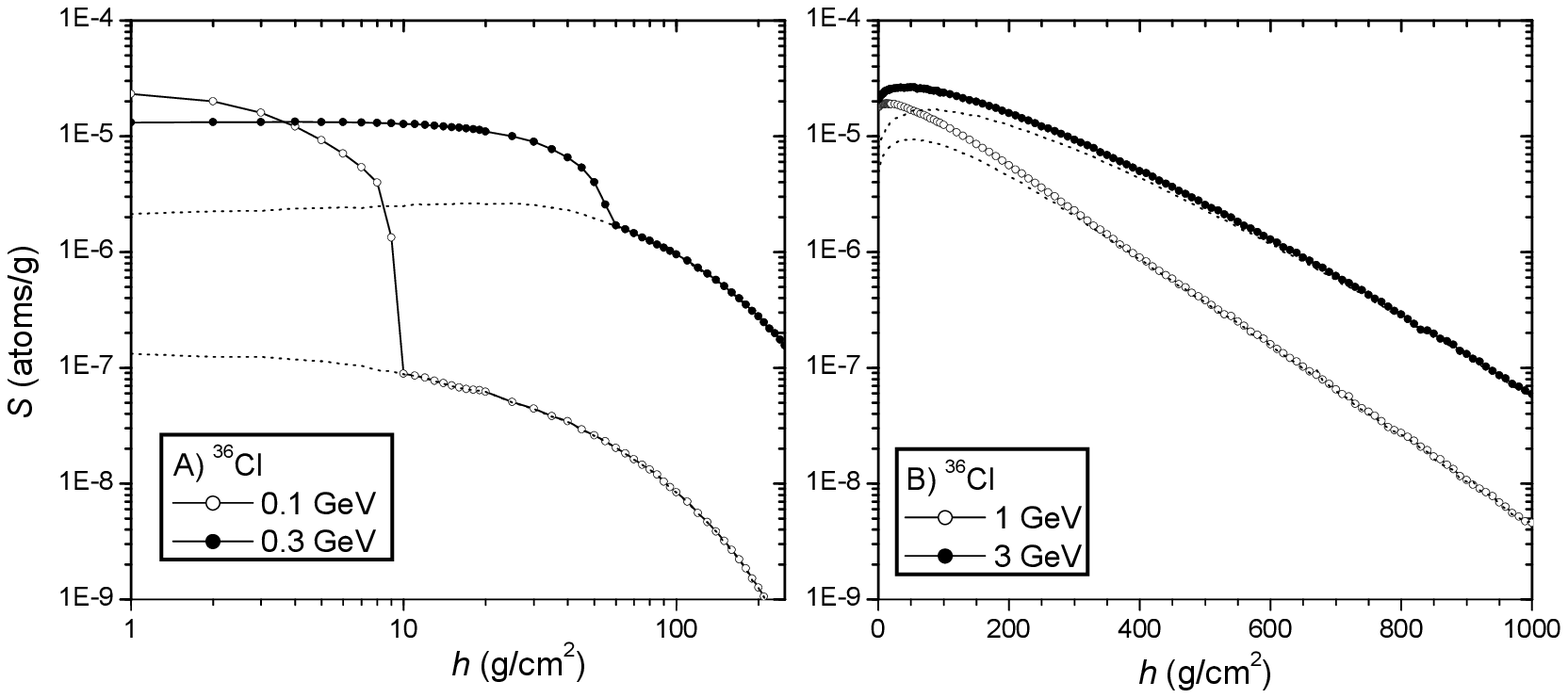}}
\caption{Production functions $S$ (see Equation~\ref{Eq:S}) of $^{36}$Cl by protons of given energy (as denoted in the legends)
 as a function of the atmospheric depth.
 Dotted lines depict contribution from secondary neutrons.
 }
\label{Fig:36Cl_depth}
\end{figure*}
Production functions computed in this way are tabulated in the Supporting information, for different isotopes and
 atmospheric depths.
We emphasize that the computational results for $\alpha-$particles are given per nucleon but not per the entire
 $\alpha-$particle.

An example of the altitude (depth) profile of the production of $^{36}$Cl is shown in Fig.~\ref{Fig:36Cl_depth}
 for primary cosmic ray protons at several selected energies ranging from 0.1 GeV to 3 GeV.
The total production is shown as solid lines with dots, while dashed curves depict contribution from secondary
 neutrons (the difference between the two is due to protons).
One can see that, for lower energy range (panel A), the production of $^{36}$Cl is dominated by
 direct spallation of atmospheric argon nuclei by primary protons in the upper atmospheric layer of a few tens of g/cm$^2$, while
 contribution of secondary particles is much smaller but becomes dominant at greater depths, where the cascade is fully developed.
This is because low-energy primary particles have insufficient energy to initiate a developed nucleonic cascade.
For energies of the primary particles greater than 1 GeV (panel B), the cascade is well developed and the
 production curve is smooth with nearly exponential attenuation with depth.

\section{Cosmogenic isotope production}

While the main result of this work, the altitude dependent yield functions, is discussed above, in this Section we present some
 applications and checks of the obtained results.
A detailed recipe on how to compute the cosmogenic isotope production at a give time and location is
 given in Appendix.

\subsection{Columnar production}

\begin{figure}
\centering \resizebox{\columnwidth}{!}{\includegraphics{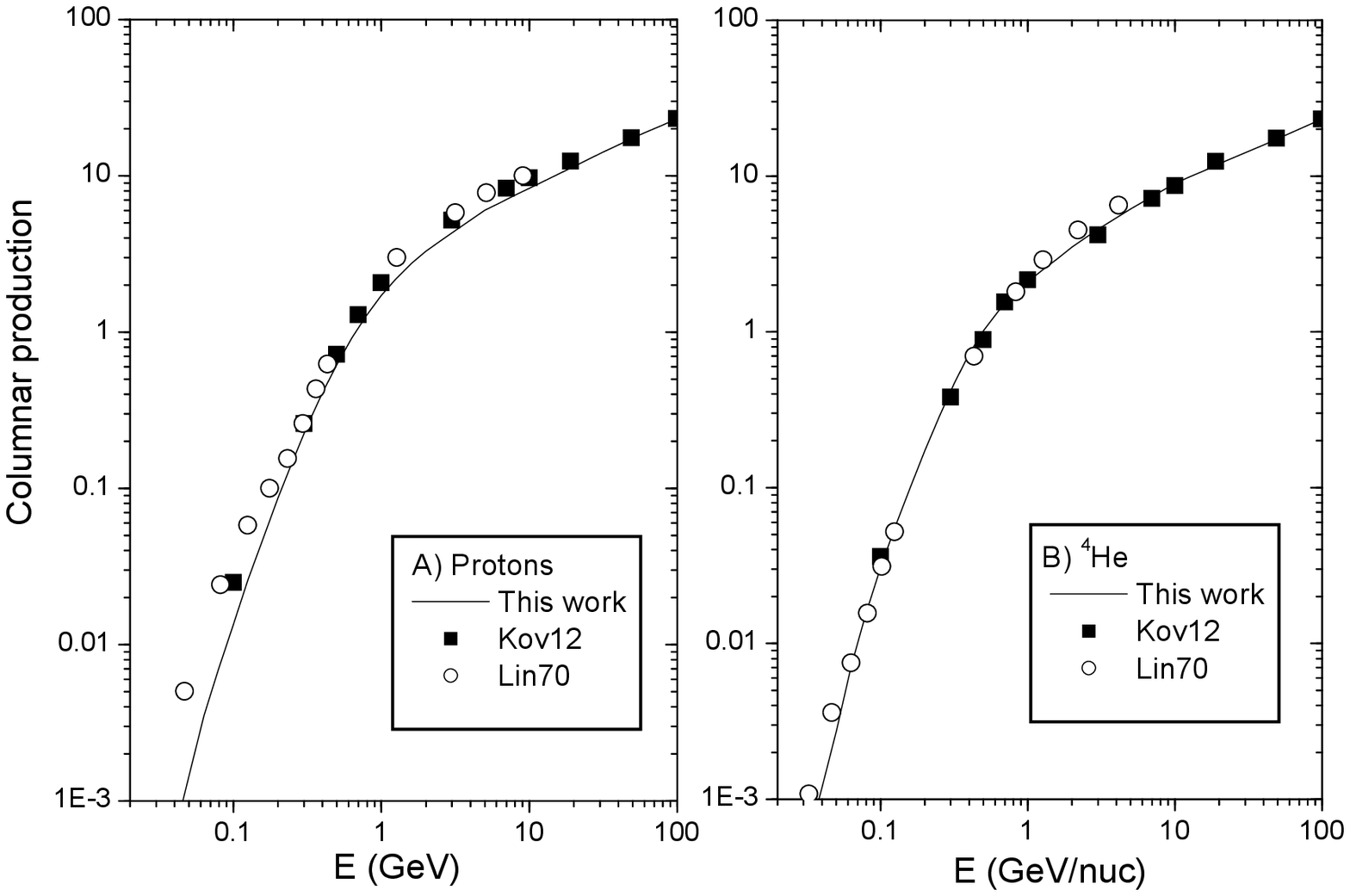}}
\caption{Columnar production $S_{\rm C}$ (atoms per incident nucleon) of the cosmogenic isotope $^{14}$C by protons (panel A)
 and $\alpha-$particles (panel B).
The black line depicts the results of this work, black squares (Kov12) and open circles (Lin70) represent
 the results by \citet{kovaltsov12} and \citet{lingenfelter70}, respectively.}
\label{Fig:C14_YF}
\end{figure}

\begin{figure}
\centering \resizebox{\columnwidth}{!}{\includegraphics{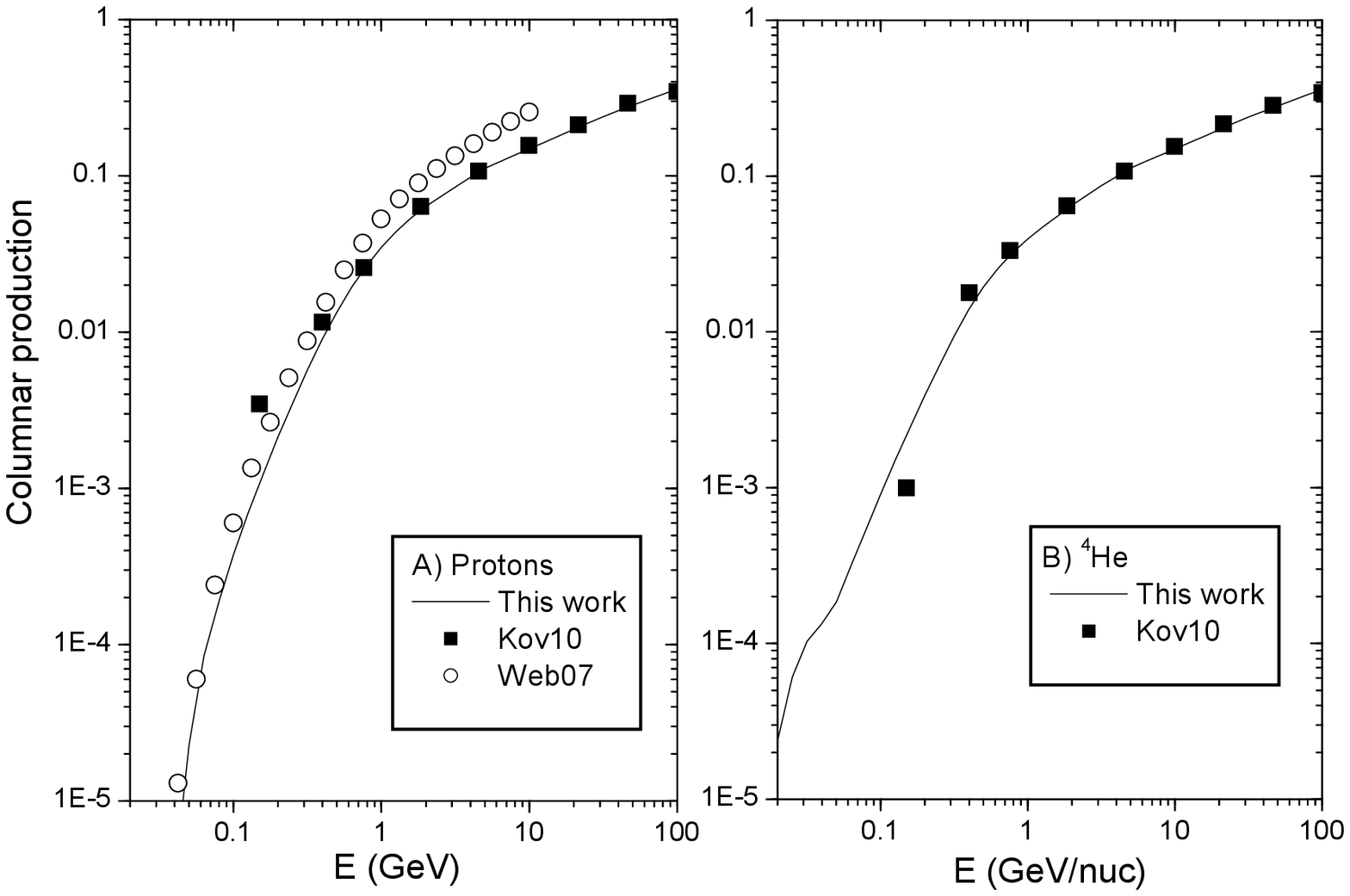}}
\caption{Columnar production $S_{\rm C}$ (atoms per incident nucleon) of the cosmogenic isotope $^{10}$Be by protons (panel A)
 and $\alpha-$particles (panel B).
The black lines depict the results of this work, black squares (Kov10) and open circles (Web07) represent
 the results by \citet{kovaltsov10} and \citet{webber07}, respectively.}
\label{Fig:Be10_YF}
\end{figure}

\begin{figure}
\centering \resizebox{\columnwidth}{!}{\includegraphics{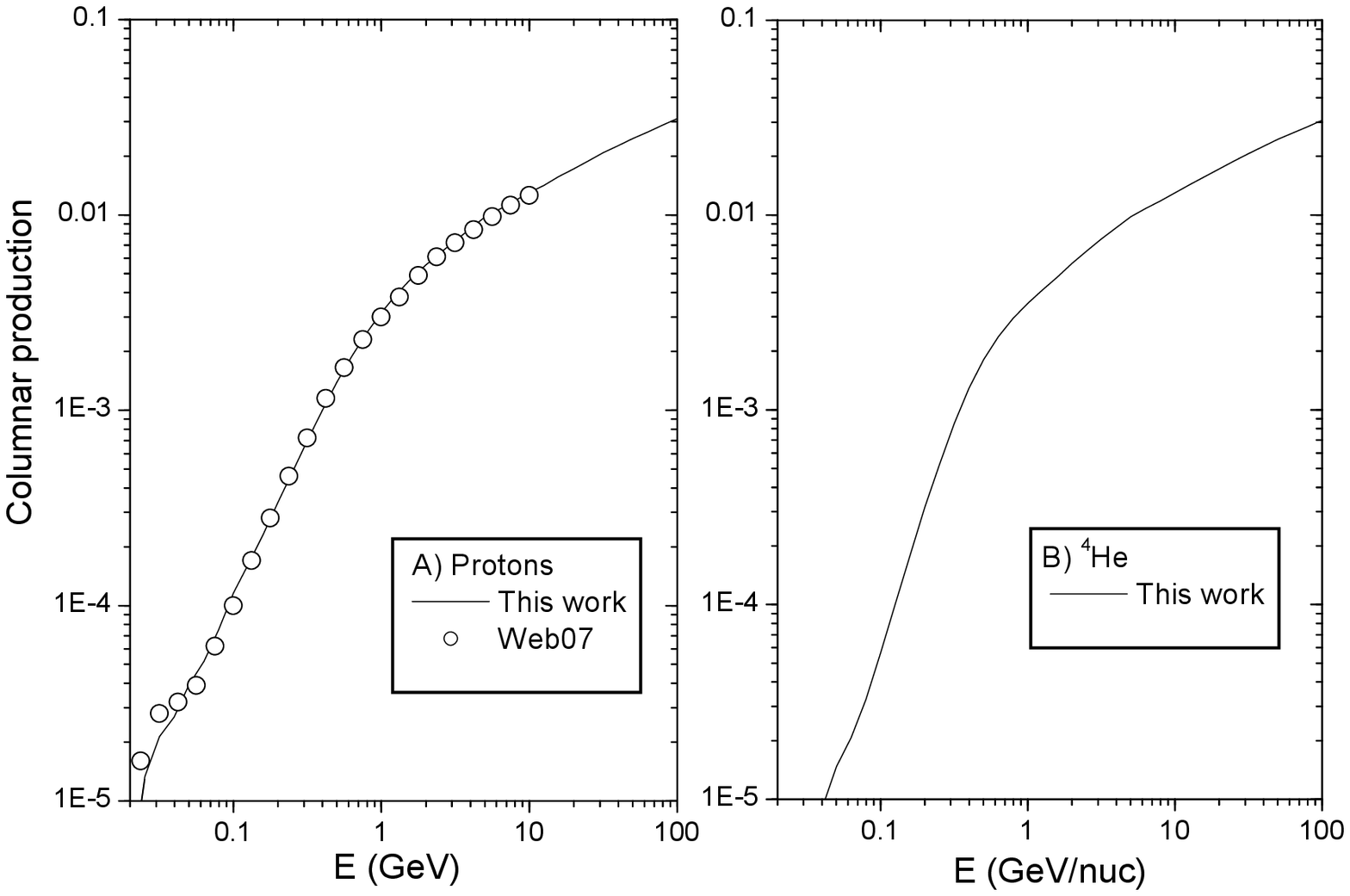}}
\caption{Columnar production $S_{\rm C}$ (atoms per incident nucleon) of the cosmogenic isotope $^{36}$Cl by protons (panel A)
 and $\alpha-$particles (panel B).
The black lines depict the results of this work, while open circles (Web07) represent
 the results by \citet{webber07}.}
\label{Fig:Cl36_YF}
\end{figure}

\begin{figure}
\centering \resizebox{\columnwidth}{!}{\includegraphics{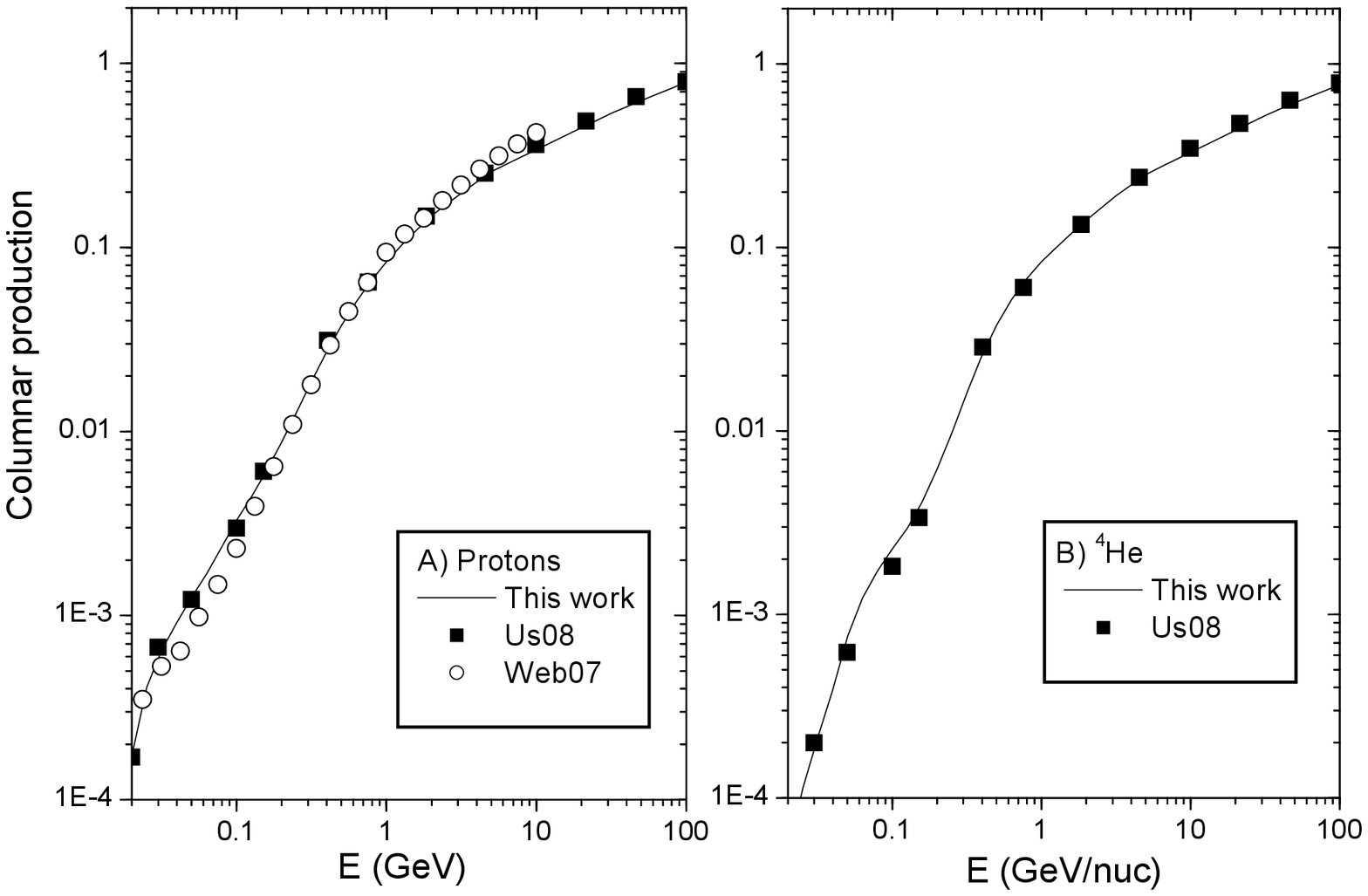}}
\caption{Columnar production $S_{\rm C}$ (atoms per incident nucleon) of the cosmogenic isotope $^{7}$Be by protons (panel A)
 and $\alpha-$particles (panel B).
The black lines depict the results of this work, black squares (Us08) and open circles (Web07) represent
 the results by \citet{usoskin_7Be_08} and \citet{webber07}, respectively.}
\label{Fig:Be7_YF}
\end{figure}

\begin{figure}
\centering \resizebox{\columnwidth}{!}{\includegraphics{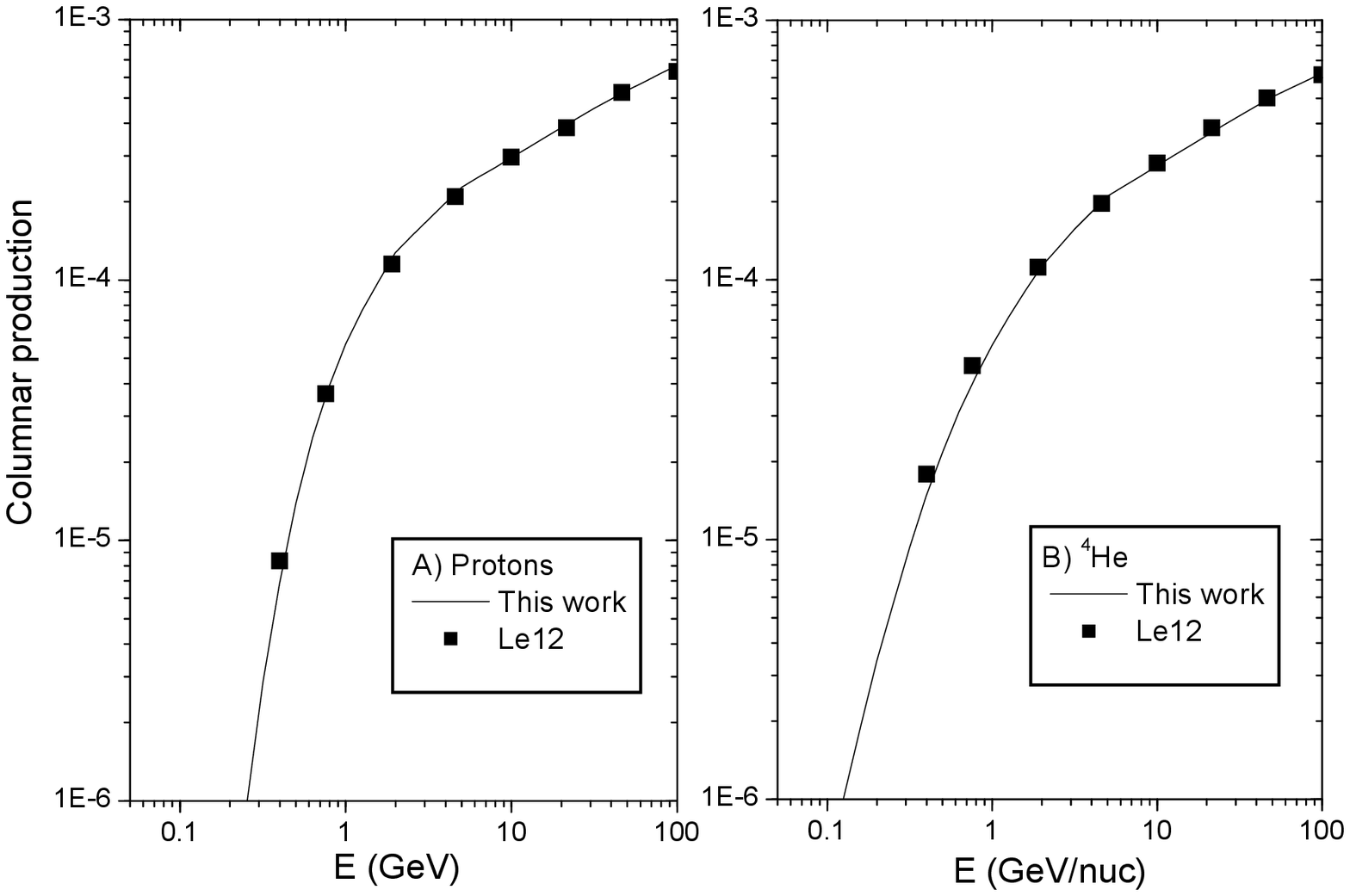}}
\caption{Columnar production $S_{\rm C}$ (atoms per incident nucleon) of the cosmogenic isotope $^{22}$Na by protons (panel A)
 and $\alpha-$particles (panel B).
The black lines depict the results of this work, while black squares (Le12) represent
 the results by \citet{leppanen12}.}
\label{Fig:Na22_YF}
\end{figure}

Columnar (viz. integrated within the entire atmospheric column) production of cosmogenic isotopes
 is often discussed \citep{webber07,kovaltsov12}:
\begin{equation}
S_{\rm C} = \int_0 ^{H} S(h)\cdot dh,
\end{equation}
where $H=1033$ g/cm$^2$ is the atmospheric depth at the mean sea level.
Accordingly, we present here columnar productions for the purpose of comparison with earlier results (see Figures \ref{Fig:C14_YF}
 through \ref{Fig:Na22_YF}).

One can see that the columnar production curves computed here are in good agreement (within 5--10\% for the energy above 100 MeV/nuc)
 with earlier results, except for one case.
While the present results for $^{10}$Be are fully consistent with those published earlier by
 \citet{kovaltsov10} for energies above 100 MeV/nuc, the discrepancy with the results by \citet{webber07} is more systematic,
 by a factor of 1.5--1.7 (Figure~\ref{Fig:Be10_YF}a).
We note that we used, for this isotope, the same cross-sections as \citet{webber03}.
We have no clear explanation for this discrepancy, especially taking into account that the result for another beryllium
 isotope $^7$Be (Fig.~\ref{Fig:Be7_YF}a), which is very similar to $^{10}$Be in the sense of production, is in good agreement
 between the two models.

We note that the columnar production is shown only for illustration, while advanced studies should include
 also complex atmospheric transport which can be modelled only using the altitude profiles of the production functions $S$.


\subsection{Global production rate}

Also, for the purpose of illustration we depict the mean global production rate, due to GCR, of the five isotopes discussed here.
The global production rate of an isotope is the averaged over the Globe columnar production rate, defined as
\begin{equation}
Q_{\rm G}(t) = {1\over 4\pi }\, \int_\Omega \int_h{Q(\phi(t),h,P_{\rm c}(\Omega,t))\cdot dh\cdot d\Omega},
\end{equation}
where $Q$ is given by Eq.~\ref{Eq:Q}, and integration is over the entire atmospheric column $h$ (as in the columnar production) and over
 the Earth's surface (longitude and latitude) $\Omega$.
The time dependence is included into variability of the modulation potential $\phi$ and in the slow changes of the
 local geomagnetic cutoff rigidity $P_{\rm c}$ at each location.

The modulation potential $\phi$ parameterizes the differential intensity of GCR in the vicinity of Earth, $J(E,\phi(t))$ (see Eq.~\ref{Eq:Q}).
It varies in accord with solar activity being low (higher GCR flux) and high (low GCR flux) around solar minima and maxima, respectively.
The formalism of the modulation potential approach is described in detail elsewhere \citep[e.g.,][]{webber03,vainio09}.
Here we use the modulation potential defined as in \citet{usoskin_bazi_11} with the local interstellar
 spectrum (LIS), according to \citet{burger00}.
We note that the exact value of $\phi$ makes sense only for a fixed reference local interstellar spectrum of GCR
 \citep{usoskin_Phi_05,herbst10}.

The large scale geomagnetic field provides additional shielding of the Earth from charged cosmic ray particles.
The shielding is often parameterized in terms of the local effective geomagnetic rigidity cutoff $P_{\rm c}$ \citep{cooke91}.
In the first approximation, the value of $P_{\rm c}$ at each location is determined by the geomagnetic latitude and
 the geomagnetic dipole moment $M$, which slowly varies on the centennial-millennial time scale in the range (6--12)$\cdot 10^{22}$ A m$^2$
 \citep[e.g.,][]{genevey08,nilsson14}.

\begin{figure}
\centering \resizebox{\columnwidth}{!}{\includegraphics{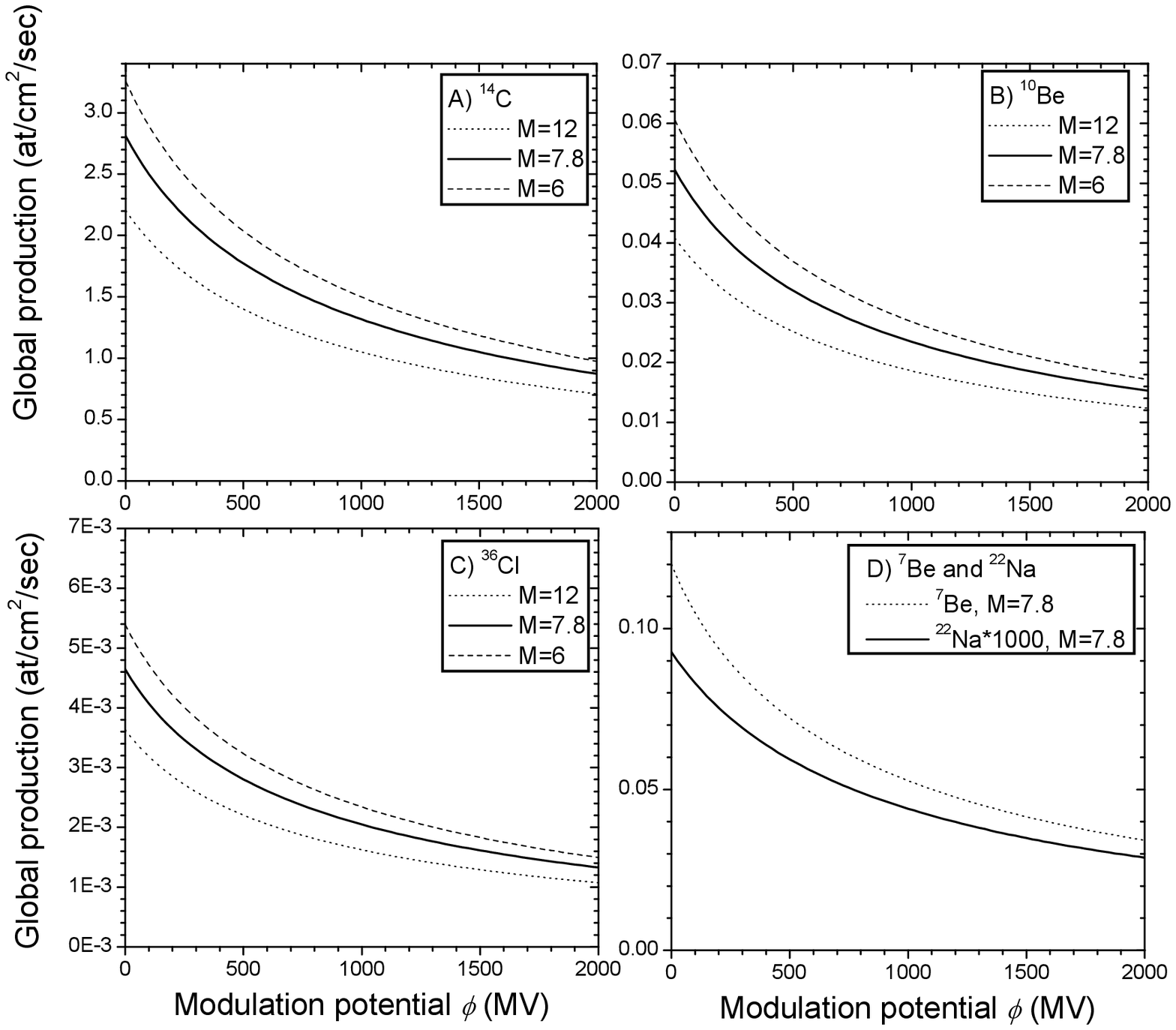}}
\caption{Global production rates of cosmogenic isotopes as a function of the modulation potential $\phi$,
 for different values of the geomagnetic dipole moment $M$ (in $10^{22}$ A m$^2$), as denoted in the legends.
 Panels A through D correspond to isotopes $^{14}$C, $^{10}$Be, $^{36}$Cl, $^{7}$Be and $^{22}$Na
 (scaled up by a factor 1000), respectively.
 }
\label{Fig:ALL}
\end{figure}

The global production rates are shown in Figure~\ref{Fig:ALL} as a function of the modulation potential $\phi$ for several values of the
 geomagnetic dipole moment $M$.
It qualitatively resembles other similar plots shown earlier \citep[e.g.,][]{masarik99,webber07}.
One can see that both parameters play an important role.
The global production rates are also shown in Table~\ref{Tab:prod} for the modern value of the geomagnetic dipole moment
 $M=7.8\cdot 10^{22}$ A m$^2$, for the cosmic ray modulation levels corresponding to the mean, minimum and maximum
 of solar cycles \citep{usoskin_bazi_11}.

We note different levels of the production rates for different isotopes: from $<10^{-4}$ for $^{22}$Na
 to a few atoms per cm$^2$ per second for $^{14}$C.
This is defined by cross-sections of the processes (neutron capture for $^{14}$C is much more likely than
 spallation reaction with a high energy threshold for other isotopes), and by abundance of the target atoms
 (argon, which is the target for $^{22}$Na, is a factor $\approx 100$ less abundant in the atmosphere than nitrogen and oxygen,
 which are targets for other elements).

\begin{table*}
\caption{Global, polar (geomagnetic pole) and equatorial (geomagnetic equator) production rates of the five cosmogenic radio-isotopes
 for the modern conditions (the geomagnetic dipole moment $M=7.8\cdot 10^{23}$ A m$^2$), for the
 mean, minimum and maximum modulation potentials: $\langle\phi\rangle =650$, $\phi_{\rm min} = 300$ and
 $\phi_{\rm max} = 1200$ MV, respectively.
 The production rates are given in atoms/cm$^2$/sec.}
\begin{small}
\begin{tabular}{c|ccc|ccc|ccc}
\hline
Isotope & \multicolumn{3}{c|}{Global production}&\multicolumn{3}{c|}{Polar production}&\multicolumn{3}{c}{Equatorial production}\\
 & mean & minimum & maximum & mean & minimum & maximum & mean & minimum & maximum\\
\hline
$^{7}$Be & $6.5\cdot 10^{-2}$ & $8.5\cdot 10^{-2}$ & $4.8\cdot 10^{-2}$ & $1.45\cdot 10^{-1}$ & $2.2\cdot 10^{-1}$ & $9.1\cdot 10^{-2}$ & $2.1\cdot 10^{-2}$ & $2.3\cdot 10^{-2}$ & $1.9\cdot 10^{-2}$\\
$^{10}$Be & $2.9\cdot 10^{-2}$ & $3.8\cdot 10^{-2}$ & $2.1\cdot 10^{-2}$ & $6.4\cdot 10^{-2}$ & $9.5\cdot 10^{-2}$ & $4.0\cdot 10^{-2}$ & $9.6\cdot 10^{-3}$ & $1.0\cdot 10^{-2}$ & $8.7\cdot 10^{-3}$\\
$^{14}$C & 1.6 & 2.07 & 1.2 & 3.42 & 5.02 & 2.21& $5.7\cdot 10^{-1}$ & $6.1\cdot 10^{-1}$ & $5.2\cdot 10^{-1}$\\
$^{22}$Na & $5.4\cdot 10^{-5}$ & $6.9\cdot 10^{-5}$ & $4.0\cdot 10^{-5}$ & $1.15\cdot 10^{-4}$ & $1.7\cdot 10^{-4}$ & $7.5\cdot 10^{-5}$& $1.8\cdot 10^{-5}$ & $1.9\cdot 10^{-5}$ & $1.6\cdot 10^{-5}$\\
$^{36}$Cl & $2.5\cdot 10^{-3}$ & $3.3\cdot 10^{-3}$ & $1.85\cdot 10^{-3}$ & $5.6\cdot 10^{-3}$ & $8.5\cdot 10^{-3}$ & $3.5\cdot 10^{-3}$& $8.3\cdot 10^{-4}$ & $8.8\cdot 10^{-4}$ & $7.5\cdot 10^{-4}$\\
\hline
\end{tabular}
\end{small}
\label{Tab:prod}
\end{table*}

Global production rates calculated here for $^7$Be, $^{10}$Be, $^{14}$C and $^{22}$Na are very close, within 5\%, to those
 published by us earlier \citep{usoskin_7Be_08,kovaltsov10,kovaltsov12,leppanen12} using an older version of the CRAC
 model.
Accordingly, comparisons with other simulation results and direct data that are discussed in great detail in those works,
 are valid also here.
The only new for us isotope is $^{36}$Cl, whose global rate agrees within 15\% with the value calculated by
 \citet{masarik09}.

We note that the mean global production makes sense only for $^{14}$C, which is globally mixed in the atmosphere.
Recent direct in-situ measurements of stratospheric radiocarbon during 2002--2005 imply the global
 production of $(2.2\pm 0.6)\cdot 10^{26}$ atoms of $^{14}$C per year \citep{kanu16}.
Our model (Fig.\ref{Fig:ALL}a) predicts the global $^{14}$C production of $2.36\cdot 10^{26}$ atoms/yr for the period
 2002--2005 (the mean $\phi=802$ MV \citep{usoskin_bazi_11}), which is in good agreement with the measurements.
Since the regional atmospheric transport and deposition prevents global mixing for other isotopes \citep{beer12}
 it makes little sense to consider the globally averaged production rate for them, especially for short-living ones such as $^{7}$Be.

For illustration, we show in Table~\ref{Tab:prod} also (geomagnetically) polar and equatorial production rates
 of the isotopes, for the modern value of the geomagnetic dipole moment.
One can see that the geomagnetic field shields cosmic rays effectively (equatorial production rates are about 15\%
 of those in polar regions), while the solar cycle variability is strongest in polar regions (a factor two vs. 10--15\%
 at the geomagnetic equator).

\section{Testing the approach}

In this Section we discuss some applications of the presented production functions for $^{10}$Be data in polar ice cores.

\subsection{Solar cycle in $^{10}$Be}

\begin{figure}
\centering \resizebox{\columnwidth}{!}{\includegraphics{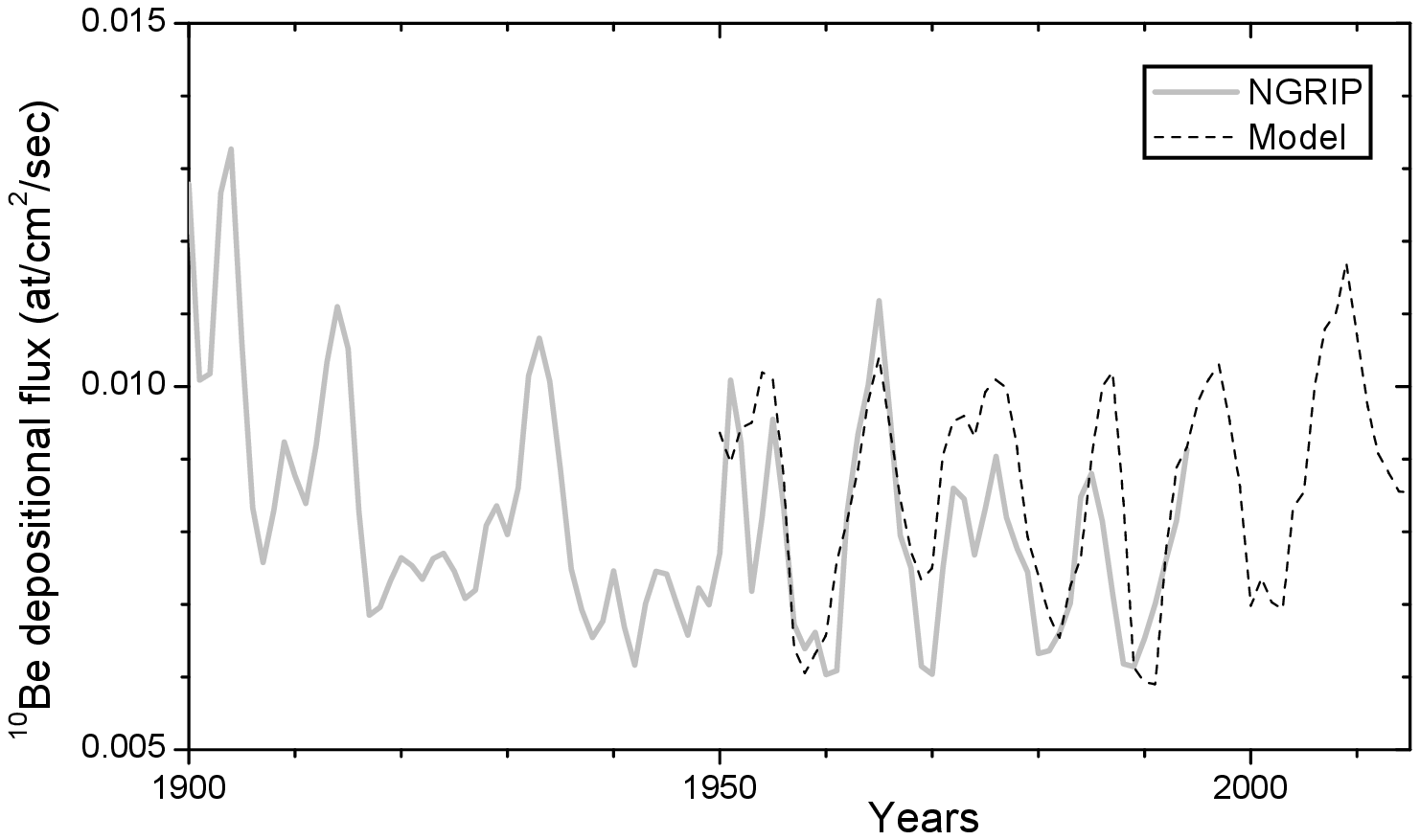}}
\caption{Annual deposition flux of $^{10}$Be in polar regions computed using the present production model
(atmospheric transport was parameterized according to \citet{heikkila09}) for the last decades.
 The grey curve depicts the measured series from the Greenland NGRIP ice core \citep{berggren09},
 while the dashed black curve is the model result.
 The data are smoothed with the 1-2-1 filter.
 }
\label{Fig:1950}
\end{figure}

As an example of an application of the approach presented here, we have computed the deposition flux
 of $^{10}$Be in the northern polar region and compared it with the measurements in the NGRIP ice core \citep{berggren09}.
The deposition flux was calculated in two steps.
First, a 3D time varying pattern of the isotope atmospheric production was calculated using the yield functions
 presented here and applying the reconstruction of the modulation potential $\phi$ based on data from
 the global neutron monitor network since 1951 \citep{usoskin_bazi_11}.
For the atmospheric transport we used a parameterization by \citet[][see Table 3 there]{heikkila09}, applying the
 mean latitudinal height profile of the tropopause.
Finally, we calculated the deposition flux of $^{10}$Be in the Northern polar region.
A 1-year delay due to the transport was applied.
The calculated $^{10}$Be flux is in very good agreement with the real data, especially for the period 1951--1970
 (see Figure~\ref{Fig:1950}).
A minor (about 5\%) discrepancy after 1970 is most likely related to the post-deposition effects in firn and/or to the regional
 climate variability on annual-decadal time scale \citep{pedro06,pedro12}.
The good agreement with data validates the yield-function for $^{10}$Be.

\subsection{The period ca. 775 AD}

\begin{figure}
\centering \resizebox{\columnwidth}{!}{\includegraphics{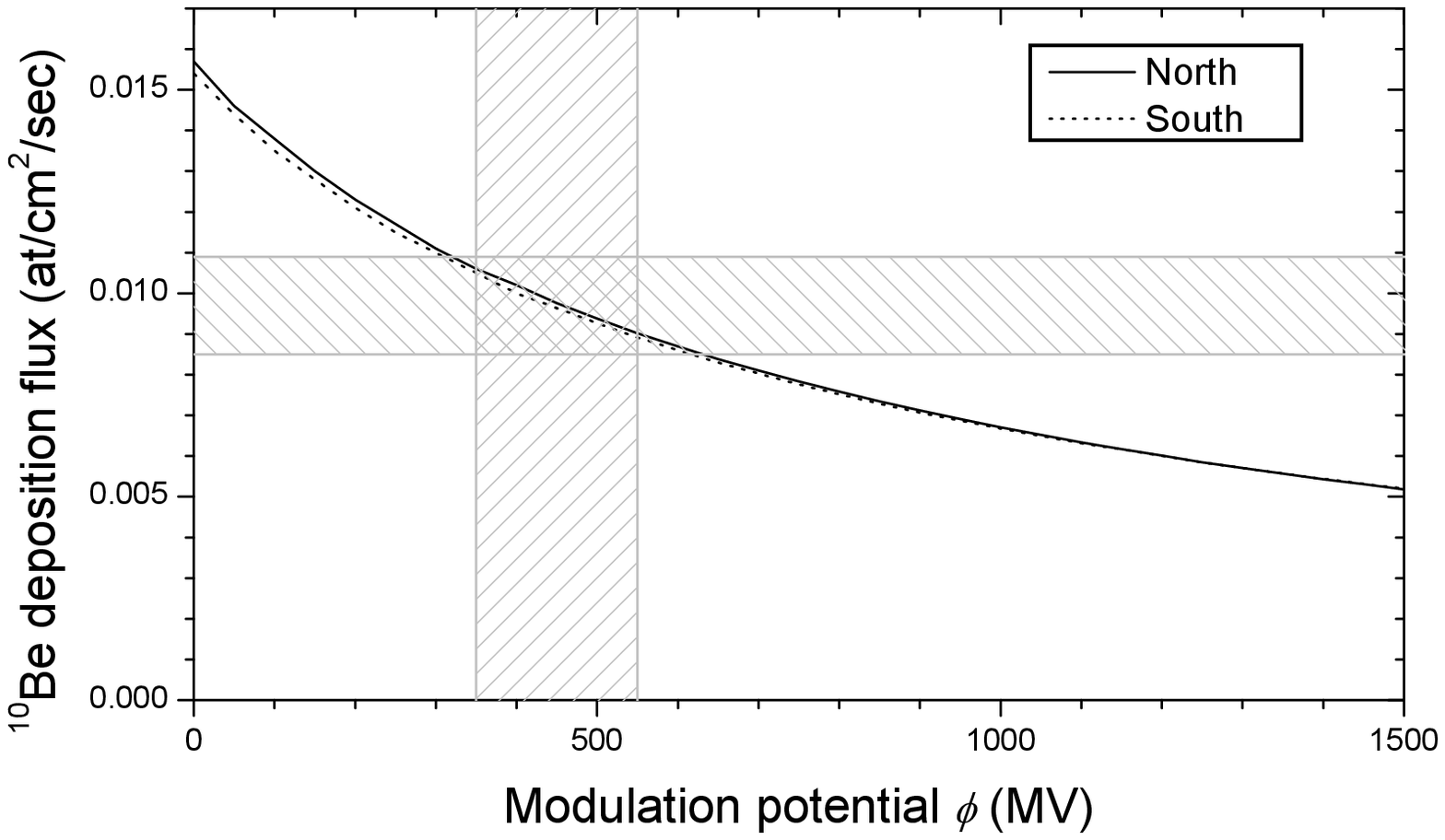}}
\caption{Deposition flux of $^{10}$Be in polar regions North and South, as denoted in the legend)
 computed using the present production model
 (atmospheric transport was parameterized according to \citet{heikkila09}) for the geomagnetic dipole moment, $M=10^{23}$ A m$^2$ as corresponding
 to the epoch of 780 AD \citep{licht13}.
 The horizontal hatched strip corresponds to the range of the decadal-mean measured fluxes of $^{10}$Be for the period 780--800 AD (with the age dating corrected)
  for four sites: Dome Fuji, Antarctica \citep{miyake15}; WDC/WAIS, Antarctica; NGRIP, Greenland, and NEEM Greenland
  \citep[for the last three sites see][]{sigl15}.
 The vertical filled grey bar represents the range of the modulation parameter $\phi$ reconstructed from $^{14}$C
  \citep{usoskin_AA_16} for the same period 780--790 AD.
 }
\label{Fig:775}
\end{figure}

The strongest known solar energetic particle (SEP) event took place in 775 AD \citep{miyake12,usoskin_775_13,mekhaldi15}.
Precise measurements of cosmogenic isotopes, in particular $^{10}$Be in ice cores, have been made for that particular period
 \citep{miyake15,sigl15,mekhaldi15}.
Not discussing the event itself, we compared the mean levels of the $^{10}$Be deposition flux for a few decades after the
 event with the prediction of our model.
In Figure~\ref{Fig:775} we compare the modelled curves for $^{10}$Be deposition flux with that measured in four polar ice
 cores in Antarctica and Greenland (see Figure caption), averaged over the period 780--800 AD (the ice core dating corrections
 applied according to \citet{sigl15}), i.e. after the event and defined by GCR, not SEPs.
These measured mean levels of the $^{10}$Be flux are shown by the horizontal hatched strip.
The vertical hatched bar corresponds to the (cycle-averaged) value of the modulation potential for that period defined independently using
 the $^{14}$C record \citep{usoskin_AA_16}.
The two black curves show the modelled deposition flux (production rate according to the present model and the atmospheric
 transport and deposition according to the parameterization by \citet{heikkila09}, the geomagnetic dipole moment $M=10^{23}$ A m$^2$
 as reconstructed for that epoch by \citet{licht13}), for the northern and southern polar regions.
One can see that the measured fluxes of $^{10}$Be are directly reproduced by the model within the possible uncertainties
 without any \textit{ad hoc} adjustment or normalization which is typically applied to $^{10}$Be data.

Accordingly, we conclude that, with the new yield function we are able to quantitatively model the production of the cosmogenic radio-isotopes
 in the atmosphere.

\section{Summary}

We have performed a new consistent and precise computation of the production of five cosmogenic isotopes,
 $^7$Be, $^{10}$Be, $^{14}$C, $^{22}$Na and $^{36}$Cl, in the Earth's atmosphere by cosmic rays.
Computations were made by means of a detailed Monte-Carlo simulation by the CRAC model
 using a recent version of the GEANT-4 tool.
The results are presented in the Supporting information in the form of tabulated yield (production) functions for a wide set of
 atmospheric depths.
We provide, for the first time, a full detailed set of the altitude profiles of the production functions
 which makes it possible to apply the results directly as input for atmospheric transport models.
Our results are in good agreement with most of the earlier published works for columnar and global isotopic production rates.
Comparison of the computations with measured data of $^{10}$Be for the last decades and also for a period around 780 AD
 validates the approach also in quantitative terms.

\section*{Appendix: Recipe for computation of the cosmogenic isotope production}

Here we present a recipe on how to compute the production rate $Q(h,P_c,t)$ of a cosmogenic isotope at a given location and time.
The location is defined by the atmospheric depth $h$ and the local geomagnetic rigidity cutoff $P_c$.

First, the yield function for a cosmic ray specie $i$ (proton or $\alpha-$particle) should be computed for the
 given atmospheric depth as $Y_i(h,E)=\pi\cdot S_i(h,E)$ (see Eq.~\ref{Eq:Y}), where $S_i(h,E)$ is taken from
 an appropriate table in the Supporting information.
Note that the energy of $\alpha-$particles should be taken as kinetic energy per nucleon.
Next, the production rate of the isotope should be computed using Equation~\ref{Eq:Q}, where the cutoff energy is
 calculated from the local geomagnetic rigidity cutoff $P_c$ using formula \ref{Eq:Tc}.
For numerical integration, values of $Y$ can be interpolated by a power-law function between the tabulated points.
The value of $P_c$ as well as spectra $J_i$ of cosmic-ray protons and $\alpha-$particles should be know independently.
For the spectra we recommend using the force-field approximation where spectra are parameterized via
 a single parameter, the modulation potential $\phi$ (see detail in \citet{usoskin_Phi_05}).
Values for the modulation potential are given, e.g., by \citet{usoskin_bazi_11} and updated at
 http://cosmicrays.oulu.fi/phi/phi.html.

\begin{acknowledgments}
This work was supported by the Center of Excellence ReSoLVE (project No. 272157).
Supporting data are included as tables in the SI file; any additional data may be obtained from IGU (email: ilya.usoskin@oulu.fi).
\end{acknowledgments}


\end{article}

\end{document}